\documentclass[final,5p,times,twocolumn]{elsarticle}
\usepackage{amssymb}
\usepackage{txfonts}
\usepackage{graphicx}

\journal{Physics Letters B}

\begin{document}

\title{One Hair Postulate for Hawking Radiation as Tunneling Process}

\author[lab1]{H. Dong}

\author[lab2]{Qing-yu Cai}
\author[lab3]{X.F. Liu}
\author[lab1]{C. P. Sun}
\ead{suncp@itp.ac.cn} \ead[url]{http://power.itp.ac.cn/~suncp/}

\address[lab1]{Institute of Theoretical Physics, Chinese Academy of Science, Beijing
100190, China}

\address[lab2]{State Key Laboratory of Magnetic Resonances and Atomic and Molecular
Physics, Wuhan Institute of Physics and Mathematics, Chinese Academy
of Sciences, Wuhan 430071, China}
\address[lab3]{Department of Mathematics, Peking University, Beijing
100871, China}

\begin{abstract}
For Hawking radiation, treated as a tunneling process, the no-hair theorem of
black hole together with the law of energy conservation is utilized to
postulate that the tunneling rate only depends on the external
qualities ( e.g ., the mass for the Schwarzschild black hole ) and
the energy of the radiated particle. This postulate is justified by
the WKB approximation for
calculating the tunneling probability. Based on this postulate, a general formula
for the tunneling probability is derived
without referring to the concrete form of black hole metric. This formula
implies an intrinsic correlation between the successive processes of
the black hole radiation of two or more particles.
It also suggests a kind of entropy conservation and thus resolves
the puzzle of black hole information loss in some sense.
\end{abstract}


\maketitle

\section{Introduction}

Hawking discovered that the black hole radiation possesses an
exactly thermal spectrum of temperature depending on the surface
gravity of the black hole~\cite{HawkingCMP1975}. Particularly, the
radiation does not depend on the details of the structure of the object
that collapsed to form the black hole. Thus, an initially pure
quantum state will evolve into a mixed thermal state as the black
hole radiates. This phenomenon, known as the paradox of black hole
information loss, obviously violates the quantum
unitarity for the closed system.

Since its appearing, many attempts~\cite{attemps} have been made to
resolve this paradox. In the previous investigations, the radiation
is always treated as possessing the thermal spectrum and the
space-time geometry is fixed. Recently, based on the WKB
approximation, the tunneling probability for the Hawking radiation
was derived in the framework of dynamical geometry. It turns out
surprisingly that the radiation spectrum is not exactly
thermal~\cite{ParikhPRL2000}. For this reason, it is found in Ref.
~\cite{ZhangPLB2009} that the successively radiated two particles
are correlated, and thus no information is lost in the radiation
~\cite{ZhangPLB2009}. Actually, by using the same approach as that
in Ref.~\cite{ParikhPRL2000}, the Hawking radiation spectra of
various black holes have been
obtained~\cite{ZhangJHEP2005,JiangPRD2006,ZhangPLB2006,ArzanoJHEP2005,BanerjeePRD2008,NozariCQG2008}
. These results verify the correlation between the successive
radiations and the conservation of the information in the radiation
~\cite{Chen2009,Zhang2009}. We find that the chain rule for the
probability is essential for the information conservation in the
black hole radiation, and we verify case by case that the chain rule
indeed holds for various Hawking radiations  coincidentally.

We observe that the above mentioned coincidence can be exactly
explained by the No-hair theorem of black hole together with the law of energy
conservation. In fact, from our ``One hair" postulate based on
the No-hair theorem and the law of energy conservation, we are able to derive a general form
of the tunneling probability of Hawking radiation without resorting to the
details of the black hole, such as its geometric structure. We are thus able to
prove that  for the tunneling probability obtained from the WKB
approximation, the chain rule is satisfied automatically and the
above mentioned coincidence is of physical necessity. It should be clear that our results demonstrate the
advantage of treating the black hole radiation as a tunneling
process.

This letter is organized as follows. In Sec.~\ref{sec:II},
Our postulate is stated based on the No-hair Theorem.  In
Sec.~\ref{sec:III}, a general formula for the tunneling
probability is derived from the postulate. In
Sec.~\ref{sec:IV},  the tunneling
rate for the Schwarzschild black hole is obtained without referring to its
geometry. In Sec.~\ref{sec:V}, the case by case verification of our postulate is given
for various black hole radiations.

\section{``One hair'' for Hawking radiation as tunneling\label{sec:II}}

It is well known that all black hole solutions of the
Einstein-Maxwell equations of gravitation and electromagnetism in
general relativity can be completely characterized by only three
externally observable classical parameters: mass, electric charge,
and angular momentum. This result is referred to as No-hair
theorem of steady black hole. For our purpose, we generalize this theorem
for the dynamic black hole as follows: the tunneling probability for the
Hawking radiation only depends on the final state of the steady
black hole and the total energy $E_{T}=E_{1}+E_{2}+...+E_{N}$
after simultaneously radiating N particles with the energies
$E_{1},E_{2},...E_{N}$. Here, there is only \textquotedblleft one hair"
quantity $E_{T}$ and the tunneling probability has nothing to do with
its partition.

\begin{figure}[tbp]
\includegraphics[bb=25 590 503 796,clip,width=9cm]{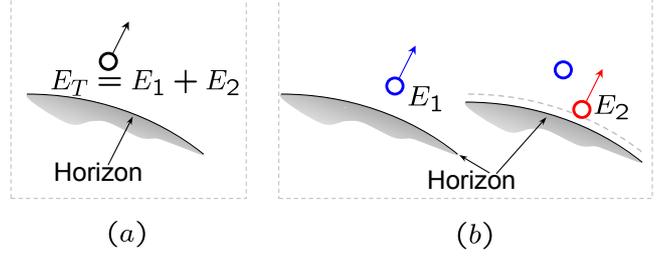}
\caption{Radiation. (a) The black hole radiates a particle with energy $E_T$%
. (b) The black hole radiates firstly a particle with energy $E_1$
and successively another particle with energy $E_2$.}
\label{Fig:radiation}
\end{figure}

To investigate the above ``One-hair " postulate, let us
consider the two processes in the Hawking radiation, illustrated in
Fig. ~\ref{Fig:radiation}:

\begin{itemize}
\item The black hole radiates a single particle with the energy $E_{\mathrm{T}}$,
as illustrated in Fig.~\ref{Fig:radiation}(a). The mass of the black
hole reduces to $M-E_{\mathrm{T}}$. The tunneling probability is
defined  as $p\left( \left\{ E_{\mathrm{T}}\right\} ;M\right) $.
The black hole can also simultaneously radiate two particles with the
energies $E_{1}$ and $\ E_{2}$ respectively. The probability of
this process is denoted by $p\left( \left\{ E_{1},E_{2}\right\}
;M\right).$ Based on the No-hair Theorem of black hole and the law of energy
conservation, we postulate the One-hair Theorem for black hole
radiation: if $E_{\mathrm{T}}=E_{1}+E_{2},$then
\begin{equation}
p\left( \left\{ E_{1},E_{2}\right\} ;M\right) =p\left( \left\{
E_{T}\right\} ;M\right).
\end{equation}

Actually, we can imagine that after the Hawking radiation the radiated
particle immediately splits into two particles with the energy $E_{x}$
and $E_{\mathrm{T}}-E_{x}$ respectively, and in the split no particular energy
partition between the two particles is preferred.
The One-hair Theorem simply means that all the splits satisfying
the law of energy conservation possess the same tunneling probability
.

\item The black hole firstly radiates a particle with the energy $E_{1}$ and
then radiates another particle with the energy $E_{2}=E_{\mathrm{T}}-E_{1}$,
as illustrated in Fig.~\ref{Fig:radiation}(b). The mass of the black
hole also reduces to $M-E_{\mathrm{T}}$. The tunneling probability for this process is
\[
p\left( \left\{ E_{1}:E_{2}\right\} ;M\right) =p\left( \left\{ E_{1}\right\}
;M\right) p\left( \left\{ E_{2}\right\} ;M-E_{1}\right)
\]
where the
conditional probability $p\left( \left\{ E_{2}\right\}
;M-E_{1}\right) $  reflects the fact that the the mass of
the black hole reduces to $M-E_{\mathrm{1}}$ after it radiats the
particle of energy $E_{\mathrm{1}}$.

We remark here that, the first radiated particle is correlated to
the second one, since the conditional tunneling probability of the
second one actually depends on the energy $E_{1}$ of the first one.
Most recently, this correlation is employed to account for the
information loss in the black hole radiation process
~\cite{ZhangPLB2009,Chen2009,Zhang2009}.
\end{itemize}

In the following we only consider the steady state of the black hole. It will take a longer time to reach the steady state than  the relaxation time of each radiation. In this case, the One-hair Theorem for black hole radiation can be
re-expressed as $p\left( \left\{ E_{1},E_{2}\right\} ;M\right)
=p\left( \left\{ E_{1}:E_{2}\right\} ;M\right) $ or
\begin{equation}
p\left( \left\{ E_{1},E_{2}\right\} ;M\right) =p\left( \left\{
E_{1}\right\} ;M\right) p\left( \left\{ E_{2}\right\}
;M-E_{1}\right).\label{eq:postualte}
\end{equation}%
Here, as only the steady solutions of the black hole radiation are
concerned, we have identified the two processes of simultaneously
and successively radiating two particles. For the multi-particle
case, we can recover the chain rule as
\begin{equation}
p\left(\left\{ E_{1}:E_{2}:...:E_{N}\right\}
;M\right)=\prod_{p}p\left(E_{p};M-\sum_{j=1}^{p-1}E_{j}\right).\nonumber
\end{equation}
based on this two-particle case. Thus, to verify the chain rule for
various Hawking radiation, we need only to prove the postulation in
Eq.~(\ref{eq:postualte}).

To justify the above observation, let us briefly review some results derived from the dynamic calculation based on the generalized
WKB approximation.  In
reference~\cite{ParikhPRL2000}, the tunneling probability for a
particle out of the black hole is defined as
\begin{equation}
p\thicksim e^{-2\mathrm{Im}S},
\end{equation}%
where $S$ is the action for an $s$-wave outgoing positive particle. The
exact form of the imaginary part of the action reads%
\begin{equation}
\mathrm{Im}S=\mathrm{Im}\intop_{M}^{M-E}\intop_{r_{\mathrm{in}}}^{r_{\mathrm{out}}}\frac{dr}{\dot{r}}dH.
\end{equation}%
Here, the Hamiltonian $H$ is defined through the radial null
geodesics equation, and particularly $H=M-E^{\prime }$ for the Schwarzschild
black hole. It is easily seen that $\mathrm{Im}S$ naturally satisfies the above stated
postulate. Then it can be concluded that the conservation of information will not be broken if
Hawking radiation is treated as tunneling process, as has been proved in many
references~\cite{ZhangPLB2009,Chen2009,Zhang2009}.

\section{Energy Dependence of Non Thermal Hawking Radiation\label{sec:III}}

In this section, we will present a derivation of the general form of the tunneling
probability  based only on the "One hair " postulate. Without losing
the generality, we assume
\[
p\left( \left\{ E\right\} ;M\right) =\exp \left[ f\left( \left\{ E\right\}
;M\right) \right] ,
\]%
where $f\left( \left\{ E\right\} ;M\right) $ is actually the
tunneling entropy for the black hole radiation. It then follows from
equation Eq.~~\ref{eq:postualte} that
\begin{equation}
f\left( \left\{ E_{T}\right\} ;M\right) =f\left( \left\{ E_{1}\right\}
;M\right) +f\left( \left\{ E_{2}\right\} ;M-E_{1}\right) .  \label{eq:entrop}
\end{equation}%
Substituting the Taylor expansion form $f\left( \left\{ \omega \right\} ;M\right)
=\sum_{n=0}A_{n}\left( M\right) \omega ^{n}$ of the function $f$ into this equation and comparing the coefficients of the terms with the same orders of  $E_{2}$, we obtain the following system of recursive equations
\begin{eqnarray*}
0 &=&A_{0}\left( M-E_{1}\right) , \\
\sum_{n=1}A_{n}\left( M\right) C_{n}^{1}E_{1}^{n-1} &=&A_{1}\left(
M-E_{1}\right) , \\
\sum_{n=2}A_{n}\left( M\right) C_{n}^{2}E_{1}^{n-2} &=&A_{2}\left(
M-E_{1}\right) , \\
&&\vdots  \\
\sum_{n=m}A_{n}\left( M\right) C_{n}^{m}E_{1}^{n-m} &=&A_{m}\left(
M-E_{1},\right)  \\
\sum_{n=m+1}A_{n}\left( M\right) C_{n}^{m+1}E_{1}^{n-\left( m+1\right) }
&=&A_{m+1}\left( M-E_{1}\right) , \\
&\vdots &.
\end{eqnarray*}%
Differentiating the left hand right hand sides of
the above equations with respect to $E_{1}$ then results in the equation
\begin{eqnarray*}
( m+1)
A_{m+1}( M-E_{1})&=&\frac{dA_{m}( M-E_{1})
}{dE_{1}}\\
&=&-\frac{dA_{m}( M-E_{1})}{dM}
\end{eqnarray*}
for each $m$.
Thus we have the recursion formula
\begin{eqnarray*}
A_{m}\left( M\right)  &=&-\frac{1}{m}\frac{d}{dM}A_{m-1}\left( M\right)  \\
&=&\frac{\left( -1\right) ^{m-1}}{m!}\frac{d^{m-1}}{dM^{m-1}}A_{1}\left(
M\right) .
\end{eqnarray*}%
and the black hole entropy can be rewritten as
\begin{equation}
f\left( \left\{ E\right\} ;M\right) =\sum_{m=1}\frac{\left(
-1\right) ^{m-1}}{m!}\frac{d^{m-1}}{dM^{m-1}}A_{1}\left( M\right)
E^{m}.
\end{equation}%
Define the entropy $G\left( M\right) $ for the black hole radiation \
through
\[
A_{1}\left( M\right) =-\frac{dG\left( M\right) }{dM},
\]%
the black hole entropy then reads
\begin{equation}
f\left( \left\{ E\right\} ;M\right) =G\left( M-E\right) -G\left( M\right) .
\label{eq:gform}
\end{equation}%
This is the main result of this paper. Obviously, $G\left( M\right)
$ in Eq. (~\ref{eq:gform}) is a conservation quantity. According to the above result, after a black hole of mass $M$ radiates a tunneling particle with energy $E$, its entropy decrease is
\begin{equation}
S\left( E,M\right) =-\ln p\left( \left\{ E\right\} ;M\right) =G\left(
M\right) -G\left( M-E\right) .
\end{equation}

In deriving the above result, it is tacitly assumed that the black hole does not carry
charge. For charged black hole a similar result can easily be obtained by the above method. In fact, when a charged black hole with charge $Q$ radiates a particle with charge $q$, the tunneling probability
can be derived as
\begin{equation}
S\left( E,q;M,Q\right) =G\left( M,Q\right) -G\left( M-E,Q-q\right) .
\end{equation}

\section{Tunneling Probability for \ Schwarzschild black hole\label{sec:IV}}

In this section, we will derive the tunneling
probability for the Hawking radiation of the Schwarzschild black
hole without referring to its dynamic geometry.

We assume that the entropy for black hole radiation is corrected
to the second order of the tunneling energy $E$, namely
\begin{equation}
f\left( \left\{ E\right\} ;M\right) =A\left( M\right) +B\left( M\right)
E+C\left( M\right) E^{2},
\end{equation}%
where $A\left( M\right) ,B\left( M\right) $ and $C\left( M\right) $
are mass-dependent functions to be determined. Then
equation (~\ref{eq:entrop}) takes the form
\begin{eqnarray*}
&&A\left( M\right) +B\left( M\right) \left( E_{1}+E_{2}\right) +C\left(
M\right) \left( E_{1}+E_{2}\right) ^{2} \\
&=&A\left( M\right) +B\left( M\right) E_{1}+C\left( M\right) E_{1}^{2} \\
&&+A\left( M-E_{1}\right) +B\left( M-E_{1}\right) E_{2}+C\left(
M-E_{1}\right) E_{2}^{2}.
\end{eqnarray*}%
gives the following equations about $A\left( M\right) ,B\left( M\right) $
and $C\left( M\right) :$
\begin{eqnarray*}
A\left( M-E_{1}\right)  &=&0, \\
B\left( M\right) -2C\left( M\right) E_{1} &=&B\left( M-E_{1}\right) , \\
C\left( M\right)  &=&C\left( M-E_{1}\right) .
\end{eqnarray*}%
It then follows that  $C\left( M\right) =k$ and $B\left( M\right) =\xi -2kM$, and the
entropy of black hole radiation is obtained as
\begin{equation}
f\left( \left\{ E\right\} ;M\right) =\left( \xi -2kM\right) E+kE^{2},
\end{equation}%
where $k$ and $\xi$ are constants. If we take $k=4\pi $ and $\xi =0$, then we recover the
well-known result by Parikh and Wilczek:
\begin{equation}
f\left( \left\{ E\right\} ;M\right) =4\pi \left[ \left( M-E\right)
^{2}-M^{2}\right] .  \label{spec}
\end{equation}%
We would like to emphasize again that, in obtaining the above result, we only make the assumption that the entropy of the
black hole is a polynomial of the radiated energy $E$, and the  details of the dynamic geometry do not come into the derivation.
If the entropy is a polynomial of $E$ of degree $1$
, then we have  $f\left( \left\{ E\right\} ;M\right)
=\xi E$ where $\xi $ is a constant independent of the mass $M$. Thus,
the conventional thermal spectrum $p^{\prime }\left( E,M\right)
=\exp \left( -8\pi EM\right) $ does not satisfy Eq.
(~\ref{eq:entrop}) about the conditional probability. In that case, $G\left(
M\right) =4\pi M^{2}=A/4$ is the usual entropy for the Schwarzschild
black hole, and is usually called Bekenstein-Hawking entropy of black hole.

According to Ref.~\cite{ZhangPLB2009}, the above spectrum function
(~\ref{spec}) indicates that the two successively radiated  particles
are actually correlated. Since Hawking radiation can carry information through this correlation between the radiated particles, the conservation of total information can be restored by taking this correlation into account.

\section{Verification of One-hair Postulate for other black holes\label{sec:V}}

In this section, we will check the radiation spectra of some well known black holes to see if they satisfy the One-hair postulate expressed by Eq.
(~\ref{eq:entrop}).

\textbf{Reissner-Nordstr\"{o}m black hole}- The tunneling
probability of a charged particle with energy $E$ and charge $q$ for the
Reissner-Nordstr\"{o}m black hole has been obtained in
Ref.~\cite{ZhangJHEP2005} as
\begin{equation}
p\left( \left\{ E,q\right\} ;M,Q\right) =\frac{\exp \left[
G_{\mathbf{RN}}\left( M-E,Q-q\right) \right] }{\exp \left[
G_{\mathbf{RN}}\left( M,Q\right) \right] },
\end{equation}%
where%
\[
G_{\mathbf{RN}}\left( M,Q\right) =\pi \left( M+\sqrt{M^{2}-Q^{2}}\right)
^{2}.
\]%
Clearly, the radiation spectrum for the \textbf{Reissner-Nordstr\"{o}m black
hole} is not thermal, and satisfies our One-hair postulate.

\textbf{Kerr black hole}-For the rotating black hole(Kerr black
hole), the tunneling probability is found in
Ref.~\cite{JiangPRD2006} as

\begin{equation}
p\left( \left\{ E\right\} ,M\right) =\exp \left[ G_{\mathbf{K}}\left(
M-E\right) -G_{\mathbf{K}}\left( M\right) \right] ,
\end{equation}%
where
\[
G_{\mathbf{K}}\left( M\right) =2\pi \left( M^{2}+M\sqrt{M^{2}-a^{2}}\right)
.
\]%
Obviously, its spectrum structure is in accordance with our One-hair postulate.

\textbf{Kerr-Newman black hole}- For the Kerr-Newman black hole, the
tunneling probability for a particle with charge $q$ is obtained in
Ref.~\cite{JiangPRD2006,ZhangPLB2006} as

\begin{equation}
p\left( \left\{ E,q\right\} ;M,Q\right) =\frac{\exp \left[
G_{\mathbf{KN}}\left( M-E,Q-q\right) \right] ,}{\exp \left[
G_{\mathbf{KN}}\left( M,Q\right) \right] ,}
\end{equation}%
where
\[
G_{\mathbf{KN}}\left( M,Q\right) =\pi \left(
M+\sqrt{M^{2}-Q^{2}-a^{2}}\right) ^{2}.
\]%
It also satisfies our postulate.

\textbf{Quantum corrected Hawking radiation}-Last, we consider the
tunneling with quantum correction for the Schwarzschild black hole.
For the quantum corrected Hawking radiation, the tunneling
probability reads
\begin{eqnarray}
p\left( \left\{ E\right\} ;M\right)  &=&\left( 1-\frac{E}{M}\right)
^{2\alpha }\exp \left[ 8\pi E\left( M-\frac{E}{2}\right) \right]   \nonumber
\\
&=&\exp \left[ G\left( M-E\right) -G\left( M\right) \right] ,
\end{eqnarray}%
where
\[
G\left( M\right) =4\pi M^{2}+2\alpha \ln M.
\]%
This tunneling probability still satisfies our postulate, thus the
information conservation is quite natural. For a detailed
discussion about the information conservation, one can refer to the
Refs.~\cite{Chen2009,Zhang2009}.

\section{Summary}

In this letter, we suggest the One-hair Postulate to describe Hawking
radiation as tunneling process based on the No-hair theorem and the
energy conservation law. This postulate for tunneling probability naturally leads to
the information conservation for the total system formed by the radiated
particles plus the remnant black hole.
Especially, this postulate is used to determine the
tunneling rate by the information (probability) theory method
rather than the dynamic geometry method. Finally, some well known
examples are presented to support the postulate. We expect the viewpoint developed in this letter will shed light on the parabox of black hole information loss.

\section*{Acknowledgement}

We thank Li You and Zhan Xu for useful discussion. The work is
supported by National Natural Science Foundation of China and the
National Fundamental Research Programs of China under Grant No.
10874091 and No. 2006CB921205.

\end{document}